\documentclass[preprint]{revtex4}
\usepackage{subeqnarray}
\usepackage{graphicx}
\usepackage{natbib}
\usepackage{amsmath}
\include{epsf} 
\begin{document}
\title{Analysis of 
Molecular Hydrogen Formation on Low Temperature Surfaces
in Temperature Programmed Desorption Experiments  
}

\author{G. Vidali$^a$, V. Pirronello$^b$, L. Li$^a$, 
J. Roser$^{a,c}$, G. Manic\'o$^b$, E. Congiu$^{a,b}$, 
H. Mehl$^d$, A. Lederhendler$^d$,
H.B. Perets$^e$, J.R. Brucato$^f$ and O. Biham$^d$}

\affiliation{a. Physics Department, Syracuse University, 
Syracuse, NY 13244, USA} 
\affiliation{b. Universit\'a di Catania, DMFCI, 
95125 Catania, Sicily, Italy} 
\affiliation{c. NASA Ames, 
Mail Stop 245-6, Moffett Field, CA, 94035, USA} 
\affiliation{d. Racah Institute of Physics, 
The Hebrew University, Jerusalem 91904, Israel}
\affiliation{e. Faculty of Physics, 
Weizmann Institue of Science, Rehovot 76100, Israel} 
\affiliation{f. INAF-Osservatorio Astronomico 
di Capodimonte, Napoli, Italy}

\begin{abstract}

The study of the formation of molecular hydrogen on low 
temperature  surfaces is of interest both because it 
enables to explore elementary steps in the heterogeneous 
catalysis of a simple molecule and due to its
applications in astrochemistry. Here we report results 
of experiments of molecular hydrogen formation on amorphous 
silicate surfaces using 
temperature-programmed desorption (TPD).  
In these experiments,
beams of H and D atoms are irradiated
on the surface of an amorphous silicate sample.
The desorption rate of HD molecules is monitored
using a mass spectrometer
during a subsequent TPD run.
The results are analyzed using
rate equations and 
the energy barriers of the processes leading to molecular hydrogen 
formation are obtained from the TPD data. 
We show that a model based on a 
single isotope provides the correct results for the activation 
energies for diffusion
and desorption of H atoms.
These results are used in order to evaluate the formation rate 
of H$_2$ on dust grains
under the actual conditions present in interstellar clouds. 
It is found that under typical conditions
in diffuse interstellar clouds, amorphous silicate grains
are efficient catalysts of H$_2$ formation when the grain
temperatures are between 9-14K. This temperature window is
within the typical range of grain temperatures in diffuse
clouds. It is thus concluded that amorphous silicates are
good candidates to be
efficient catalysts of H$_2$ formation in diffuse clouds.

\end{abstract}

\maketitle

\section{Introduction}
\label{sec:introduction}

Few are the studies of the formation of molecular 
hydrogen on low temperature surfaces. 
One of the pioneering experiments was done
in the 1970's 
by the group of Giacinto Scoles, 
who measured the scattering, sticking 
and energy deposition of atomic and molecular hydrogen beams  
on the surface of bolometers (semiconductor thin films) 
at liquid helium temperature 
\cite{Marenco1972,Schutte1976,Govers1980}.
It was found that both the sticking coefficient and the hydrogen 
recombination rate depend on the coverage of H$_2$ on the target surface.
It was also shown  
that the heat released in  the formation of molecular hydrogen 
causes the desorption of hydrogen molecules that have been pre-adsorbed  
on the surface. Thus, a molecule just formed is immediately ejected 
from the surface. 
These experiments offered a rare view of the
interaction of hydrogen atoms and molecules in the physical 
adsorption regime and 
a connection 
with processes in interstellar space.
However, the
sample temperature of 3-4 (K) was well below that of 
interstellar dust grains, 
the coverage of the sample with atoms/molecules 
was high and the ice layer
not fully characterized. 
These conditions made it difficult to obtain a quantitative
understanding of actual gas-dust grain processes 
in astrophysical environments. 

Molecular hydrogen (H$_2$), the most abundant molecule in the Universe, 
influences the chemical make-up of the Cosmos 
\cite{Duley1984,Williams1998}
and is instrumental in the formation of stars by contributing 
to the cooling during the gravitational collapse of molecular clouds. 
The challenge of explaining the formation 
of molecular hydrogen in space begins with the 
realization that the stabilization of the nascent molecule in 
the bonding of two (neutral) hydrogen (H) atoms involves the forbidden 
transition to the ground state. 
Three-body gas-phase interactions 
are too rare to contribute significantly to H$_2$ formation
in cold clouds
\cite{Duley1984},
but may take place in other environments such as interstellar shocks.
Under conditions observed in interstellar clouds, other gas-phase routes 
(such as: H$+e \rightarrow$ H$^-$ + $h\nu$, 
H$^-$ + H $\rightarrow$ H$_2$ + $e$; 
or, less frequently, 
H+H$^+$ $\rightarrow$ H$_2^+$, H$_2^+$ + H $\rightarrow$ H$_2$ + H$^+$) 
do not  make 
enough H$_2$ to counterbalance the known destruction rate due to 
UV photons.  
\cite{Duley1984}. 

In the 1960's, Salpeter and collaborators proposed a model in 
which H$_2$ formation occurs on the surfaces of interstellar
dust grains
\cite{Gould1963,Hollenbach1970,Hollenbach1971a,Hollenbach1971b}.
These grains are formed in the envelopes of massive late-stars 
and in novae and supernova explosions. 
They are made of carbonaceous materials and of silicates.  
Their sizes 
exhibit a broad power-law like distribution
between 1-100 nm
\cite{Mathis1977,Draine1984,Weingartner2001}.
Observations of scattered, absorbed and emitted starlight, and 
laboratory work, show that in the interstellar medium (ISM), 
silicate grains are amorphous and mostly of composition 
(Fe$_x$Mg$_{1-x}$)$_2$SiO$_4$,
where
$0<x<1$ 
\cite{Tielens2005}.
There is, on average, one dust grain per about
10$^{12}$ hydrogen atoms, 
and the grains account for about 1\% of the mass of
interstellar clouds.
Kinematic calculations show that 
in order to produce enough molecular hydrogen to 
counterbalance the destruction rate, 
the catalysis on grain surfaces must be efficient.
More specifically, 
the processes of hydrogen sticking, 
migration and bond formation on the grains 
must convert at least about 
$\sim 30 \%$ 
of the adsorbed hydrogen atoms into molecular form
\cite{Hollenbach1970}.

With some exceptions
\cite{Hasegawa1992,Herbst2003},
chemical models that look at the chemical evolution of an 
interstellar cloud, have largely ignored or underplayed 
the coupling of gas and dust. 
However, 
observations, experiments and calculations 
are pointing to the fact 
that the formation of key ISM molecules 
[such as H$_2$, formaldehyde (H$_2$CO), and methanol (CH$_3$OH)] 
takes place on dust grains,
as gas-phase 
reactions are too slow in these particular cases
\cite{Geppert2006,Hidaka2004,Watanabe2005,Garrod2006}. 
As far as the formation of molecular formation is concerned, 
there is a great need to 
know the basic 
mechanisms of reaction 
(Langmuir-Hinshelwood, Eley-Rideal or hot atom),  
characteristic energies for various processes 
(diffusion and desorption) 
and kinetic parameters 
of dust-catalyzed reactions so they can be 
used in models of interstellar chemistry.

It is within this framework that in the late 1990's we 
began a series of investigations on the formation of 
molecular hydrogen on analogues of dust grains 
\cite{Pirronello1997a,Pirronello1997b,Pirronello1999}.
These experiments, inspired by Scoles' work, were aimed at 
 combining
tools of surface science, chemical physics and 
low temperature physics in order to recreate the  
environmental conditions of the interstellar space
and overcome some of the limitations of prior experiments, such as: 
high fluxes of H, too low sample temperatures
and not adequately 
characterized materials (For a review of early experiments, 
see Ref. \cite{Vidali1998}).
In practice, the experiments have to be done at  
low background pressure, low sample 
temperatures and low fluxes of atoms impinging on the samples. 
The first two requirements are relatively easily achieved. 
Even taking   special care to obtain fluxes of low energy (200-300K) 
hydrogen atoms, it is not possible to either produce or detect
as low fluxes of atoms as appear in the ISM. 
Thus,  carefully designed 
theoretical and computational tools need to be used 
to simulate the actual processes 
occurring is the ISM using the results of the experiments.

The formation of molecular hydrogen on surfaces has 
been explored at length in the past, but most of the 
work has been on characterized surfaces of metals and 
semiconductors, and at much higher surface temperatures 
and fluxes (or coverages) 
than in the regime we are interested in.  
On low temperature surfaces, efficient recombination can 
occur only if the mobility of hydrogen is high. 
There are situations when this does not have to be verified, 
as in the Eley-Rideal and hot atom mechanisms, in which 
H atoms from the gas phase directly interact with the 
target hydrogen atoms or move on the surface at 
superthermal energy. 
Such mechanisms have been 
shown to be working in the interaction of H with 
H-plated metal
\cite{Rettner1994,Eilmsteiner1996}, 
silicon 
\cite{Khan2007}
and graphite 
\cite{Zecho2002}
surfaces. 
Although there are certain  interstellar 
environments where these mechanisms enter into play, 
the diffuse cloud environment - where the coverage of H 
atoms on a grain at any given time is very low - 
is not one of them. 
Thus, we expect that H atoms 
in our experiments to experience physical adsorption forces 
and the dominant mechanism of reaction is expected to be the 
Langmuir-Hinshelwood reaction.

In the experiments, the sample is exposed to well 
collimated beams of hydrogen (H) and deuterium (D)
atoms. 
The production of HD molecules occurring on the surface of a dust grain
analogue is measured both during the irradiation with the beams and
during a subsequent temperature programmed desorption (TPD) experiment. 
In order to disentangle the process of diffusion from the one of
desorption, additional experiments 
are carried out in which molecular species
are irradiated on the sample and then are induced to desorb.

We first studied the formation of  hydrogen deuteride on a 
telluric polycrystalline sample of  olivine
\cite{Pirronello1997a,Pirronello1997b}. 
This was followed by studies of HD formation on amorphous 
carbon and amorphous water ice 
\cite{Pirronello1999,Roser2002,Roser2003}.
Water ice is known 
[together with other condensables, such as 
carbon dioxide (CO$_2$), carbon monoxide (CO), and CH$_3$OH] 
to coat grains in molecular clouds, 
under conditions of high density, low temperature
(10-15K) and in the absence of UV radiation.
The high density is important for effective mantle formation,
while the low temperature is essential in order to
avoid mantle evaporation.
Thus, shielding is required from both UV radiation
(which causes photodissociation)
and from thermal heating sources, namely stars and protostars.
%where the self-shielding of H$_2$ protects the inner part 
%of the cloud from UV photons, thus keeping grains at low 
%temperatures ($\sim 10-15$K). 
Other groups studied 
the formation of molecular hydrogen on amorphous water ice surfaces
\cite{Hornekaer2003}
as well as the desorption of H$_2$
from these surfaces
\cite{Hornekaer2005,Dulieu2005,Amiaud2006}.
% Write about results of other groups on ice
% porosity, well depths dist.
% Related to comment 3

Measurements of the kinetic energy of hydrogen
molecules emerging from amorphous water ice 
show that the molecules have nearly thermal energies
\cite{Roser2003,Hornekaer2003}.
The ro-vibrational states of excitation of the just-formed molecules 
leaving the surface were also studied using graphite samples
\cite{Hornekaer2003,Creighan2006}.
For a review of recent experimental work on the formation
of molecules on astrophysically relevant surfaces see Ref.
\cite{Williams2007}.

The picture that emerges from the experimental 
studies on H$_2$ formation is the following. 
Molecular hydrogen formation takes place via the 
Langmuir-Hinshelwood mechanism, at least in the
range of surface temperatures 
(during irradiation)
that was sampled (5-16K).
The results of our TPD experiments
were analyzed
using rate equation models 
\cite{Katz1999,Cazaux2002,Cazaux2004,Perets2005}.
In this analysis
the parameters of the rate equations
were fitted to the experimental TPD curves.
These parameters include
the energy barriers for atomic
hydrogen diffusion and desorption and 
the energy barrier for molecular hydrogen
desorption. 
Using the values of the
parameters that fit best the experimental results, 
the efficiency of
hydrogen recombination on the polycrystalline 
olivine, amorphous carbon and water ice surfaces 
was calculated
for interstellar conditions.
By varying the temperature and flux over 
the astrophysically relevant range, 
the domain in which 
there is non-negligible recombination efficiency
was identified. 
It  was found that the
recombination efficiency is highly temperature dependent. 
For each of the samples, 
there is a narrow window of high efficiency 
along the temperature axis,
which slowly shifts to higher temperatures 
as the flux is increased. 
Contrary to expectations, for astrophysically relevant fluxes, the
formation of H$_2$ on polycrystalline olivine occurs in a 
temperature range which is too low
and too narrow for making molecular hydrogen but 
in very selected environments.
On other surfaces, such as amorphous carbon and 
amorphous water ice, efficient molecular hydrogen
formation occurs on a wider sample temperature range, 
which is within the typical range of
grain temperatures in diffuse and dense clouds 
\cite{Katz1999,Cazaux2002,Cazaux2004,Perets2005}.

Since silicates make up a significant fraction of the interstellar dust, 
and given our results that pointed at a less than optimal efficiency 
of polycrystalline olivine in catalyzing the formation of molecular 
hydrogen, we decided to revisit the formation of H$_2$ on such a 
class of materials.
Although crystalline silicates 
are observed in circumstellar envelopes, in the ISM silicates are 
mostly amorphous.   
We thus studied the formation of molecular hydrogen 
on amorphous silicate samples, 
(Fe$_{x}$, Mg$_{1-x}$)$_{2}$SiO$_{4}$, 
$x=0.5$,
produced by laser ablation
(with wavelength of 266 nm) of a mixed MgO, FeO and SiO$_{2}$ target in an
oxygen atmosphere (10 mbar). 
The optical and stochiometric characterization
of the samples produced with this technique is given elsewhere 
\cite{Brucato2002}. 
Measurements of the amorphous silicate samples 
by scanning electron microscope show rough surfaces 
with droplets of different sizes.
However, unlike the amorphous water ice,
no indications of porosity were observed.
A brief report on the formation of molecular hydrogen 
on an amorphous sample of composition 
(Fe$_{0.5}$Mg$_{0.5}$)$_2$ SiO$_4$ 
at low irradiation temperature (around 5K) 
appears in Ref.
\cite{Perets2007}.

In this paper we present data on the formation of HD at a 
higher irradiation temperature (10 K) which is at the lower edge of the 
range of the dust temperature in the ISM. 
We describe in detail the 
methodology for analyzing the results
of the TPD experiments. 
We show that the results,
based on measurements of HD formation
are applicable for the evaluation of H$_2$ formation
rate in interstellar clouds.
Results from surfaces of samples with other compositions 
(Fe$_x$,Mg$_{1-x}$)$_2$SiO$_4$, 
$0<x<1$, 
will be given elsewhere.

The paper is organized as follows: 
In Sec.
\ref{sec:experimental} 
we present a review of the experimental methods. 
TPD results for HD formation on an 
amorphous silicate surface are given in 
Sec.
\ref{sec:results}.
The rate equation model is presented in Sec. 
\ref{sec:model}
and
is used in Sec.  
\ref{sec:analysis}
for the analysis of the experimental results.
Applications 
to interstellar chemistry
are considered
in Sec. 
\ref{sec:applications}.

\section{Review of Experimental Methods}
\label{sec:experimental}

The apparatus consists of two atomic/molecular beam lines 
aimed at a target located in a 
ultra-high vacuum (UHV) chamber. 
The triple 
differentially pumped beam lines have each a 
radio-frequency (RF) dissociation source; 
typical dissociation rates are in the 75-90$\%$ range and are 
measured when the beams enter the sample chamber. 
The atoms can be cooled by passing them through a 
short aluminum nozzle connected to a liquid nitrogen 
reservoir via copper braids. 
The fluxes $F_{\rm H}$ and $F_{\rm D}$, of H and D atoms,
respectively, are both estimated to be equal to
$F_0 = 10^{12}$ 
(atoms cm$^{-2}$s$^{-1}$)
\cite{Roser2002}.

The detector, located in the main UHV chamber, 
is a differentially pumped quadrupole mass spectrometer 
that can be rotated around the sample; it is used to 
measure the signals proportional to the number of 
particles in the beams and of the molecules evolving 
from the surface either during the irradiation phase 
or during the thermal desorption.
The sample is mounted on a liquid-helium cooled 
sample holder with thermal shields. 
The sample can be rotated around its axis. 
A heater in the back of the sample can heat 
it to 400K for cleaning. 

A typical experiment proceeds as follows.
During the baking of the apparatus and prior each 
series of data taking, the sample temperature is taken 
to $\sim 400$K. After cooldown, when the desired 
sample temperature is reached, the sample is 
exposed to two converging beams of H and D 
(prior the pumping down, laser beams are shone 
through the beam lines to make sure they point 
at the same spot on the sample). 
Using two isotopes is essential in order that 
the fraction of 
undissociated molecules and the background 
pressure of H$_2$ will not affect our measurements. 
During the irradiation phase of the experiment,
as well as the subsequent TPD phase, 
the detector monitors the increase of HD partial 
pressure due to HD formation on and release 
from the surface. 
After irradiation time of $t_0$ (s),
the beams are turned off and the sample temperature is raised,  
either by shutting 
off the flow of liquid helium or by increasing the power to the heater. 
The sample temperature
vs. time, $T(t)$,
is measured by an iron-gold/chromel thermocouple and a calibrated
silicon diode placed in contact with the sample. 
The temperature curves are nonlinear but highly reproducible.
To account for this nonlinearity, 
the TPD curves in Figs. 
\ref{fig:2}-\ref{fig:4}
show the instantaneous HD desorption rate
vs. temperature rather than vs. time. 
The heating rate is steep in the beginning and 
gradually decreases.
The temperature curves for the experiments
analyzed in this paper are shown in the insets
in Figs. 
\ref{fig:1}
and 
\ref{fig:2}.
The typical irradiation times 
are 1 to 2 minutes (although, the actual time the 
sample is exposed to the beam is half that, 
since the beams are chopped with a mechanical 
selector of 50\% duty cycle). 
The typical coverages, assuming a sticking 
coefficient of 1 are of the order of a few percents of a 
layer. 

During the experiment, it is checked that 
after a thermal desorption there is no re-adsorption 
of HD on the surface. 
Also, in the experiments done 
at the lowest sample temperatures during irradiation (such as 5.6K), 
care is taken to make sure 
that the sample remains in this low temperature long enough
to reach
thermal equilibration (as judged by the repeatability 
and quality of the results). Further details can be 
found in other publications
\cite{Vidali2006}.

\section{Experimental Results}
\label{sec:results}

Here we report TPD results for HD desorption 
from amorphous silicates  
of composition (Fe$_x$,Mg$_{1-x}$)$_2$SiO$_4$, 
where $x=0.5$,
after irradiation by H+D beams at a sample temperature of 10K.
Three specimen were used, yielding comparable results.  
These results complement and extend those reported
in Ref.
\cite{Perets2007},
where we analyzed TPD results for irradiation of H and D
(hereafter H+D) 
at a low sample temperature ($\sim$5K) and in which 
the exposure was varied.
In these experiments we find that most of 
the molecules that form on the surface come off 
during the TPD; therefore, we focus
our analysis on this aspect of the experiment.
In a separate set of experiments, beams of HD molecules
were irradiated on the same surface.

\begin{figure} 
\includegraphics[scale=0.55,angle=270]{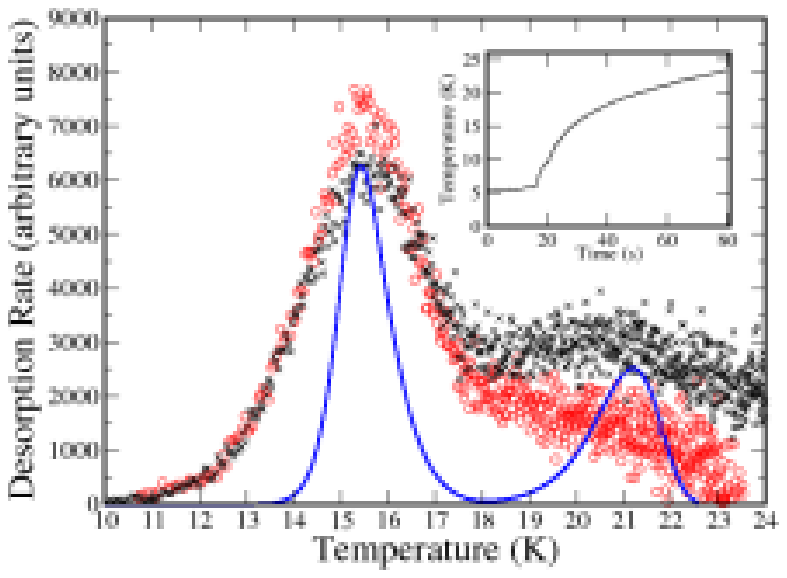}
\caption{
TPD curves of HD desorption after irradiation 
H+D atoms (circles) and HD molecules ($\times$)
on the amorphous silicate sample at surface
temperature of 5.6K. 
The fit 
of the TPD curve for H+D irradiation,
obtained using the rate equations, is also
shown (solid line).
The temperature ramp $T(t)$ during the TPD stage
is shown in the inset. 
}
\label{fig:1}
\end{figure}

\begin{figure} 
\includegraphics[scale=0.55,angle=270]{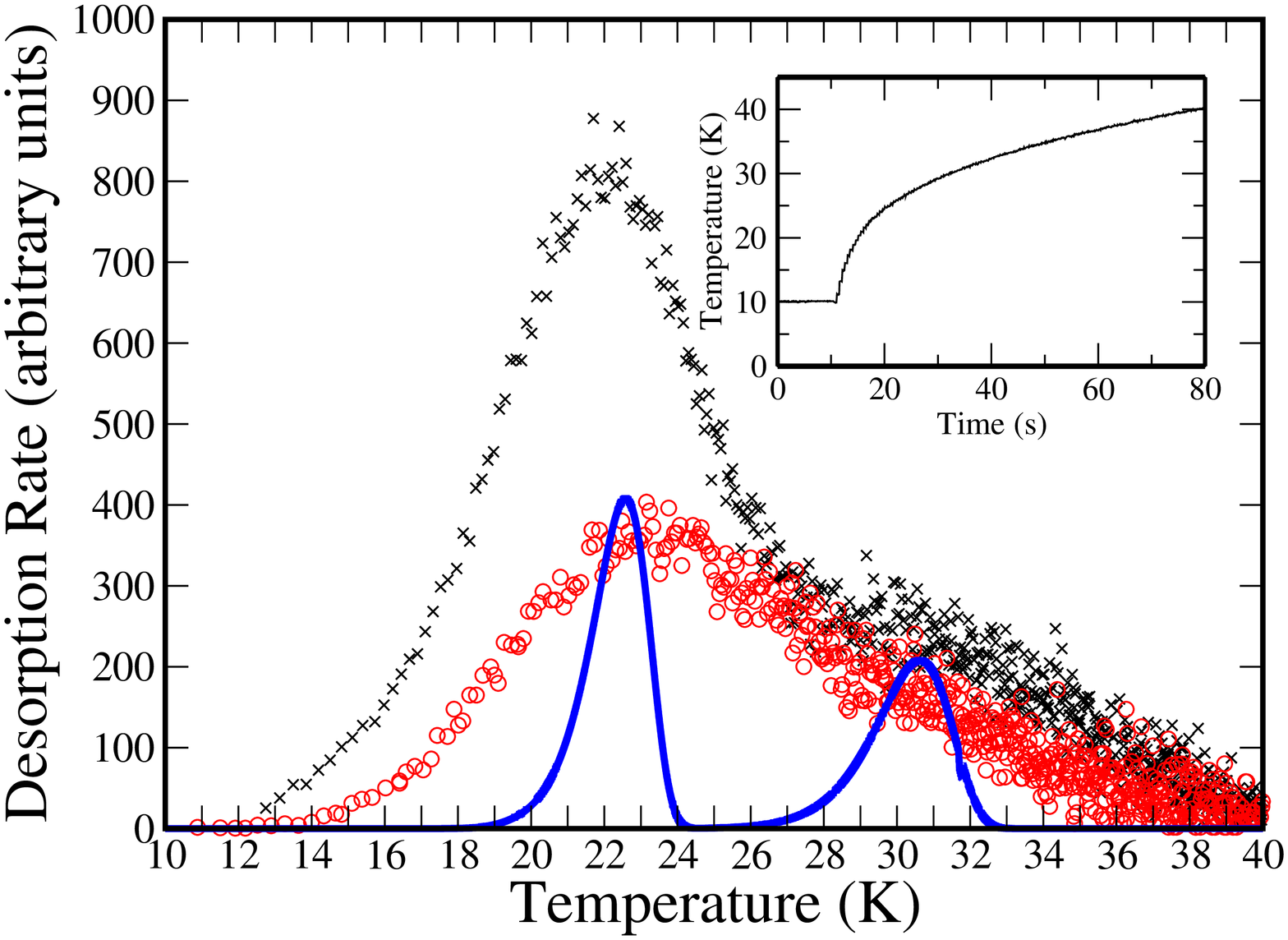}
\caption{
TPD curves of HD desorption after irradiation 
H+D atoms (circles) and HD molecules ($\times$)
on the amorphous silicate sample at surface
temperature of 10K. 
The fit 
of the TPD curve for H+D irradiation,
obtained using the rate equations, is also
shown (solid line).
The temperature ramp $T(t)$ during the TPD stage
is shown in the inset. 
}
\label{fig:2}
\end{figure}

\begin{figure} 
\includegraphics[scale=0.55,angle=270]{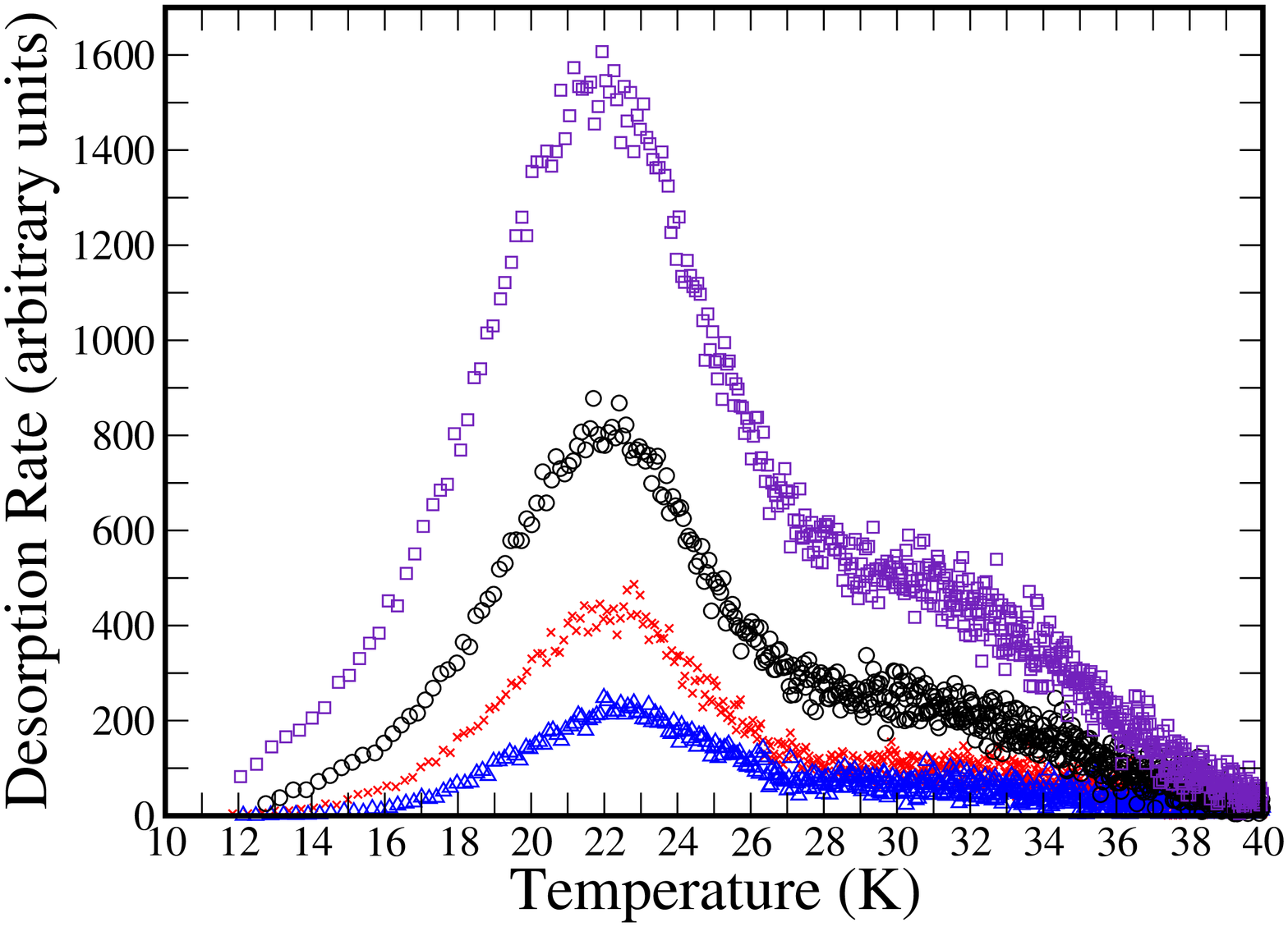}
\caption{
TPD traces obtained after
irradiation with HD molecules at a sample temperature
of 10K. The irradiation times are
30 (triangles), 60 ($\times$), 120 (circles) and 240 (squares) seconds.
}
\label{fig:3}
\end{figure}

\begin{figure} 
\includegraphics[scale=0.55,angle=270]{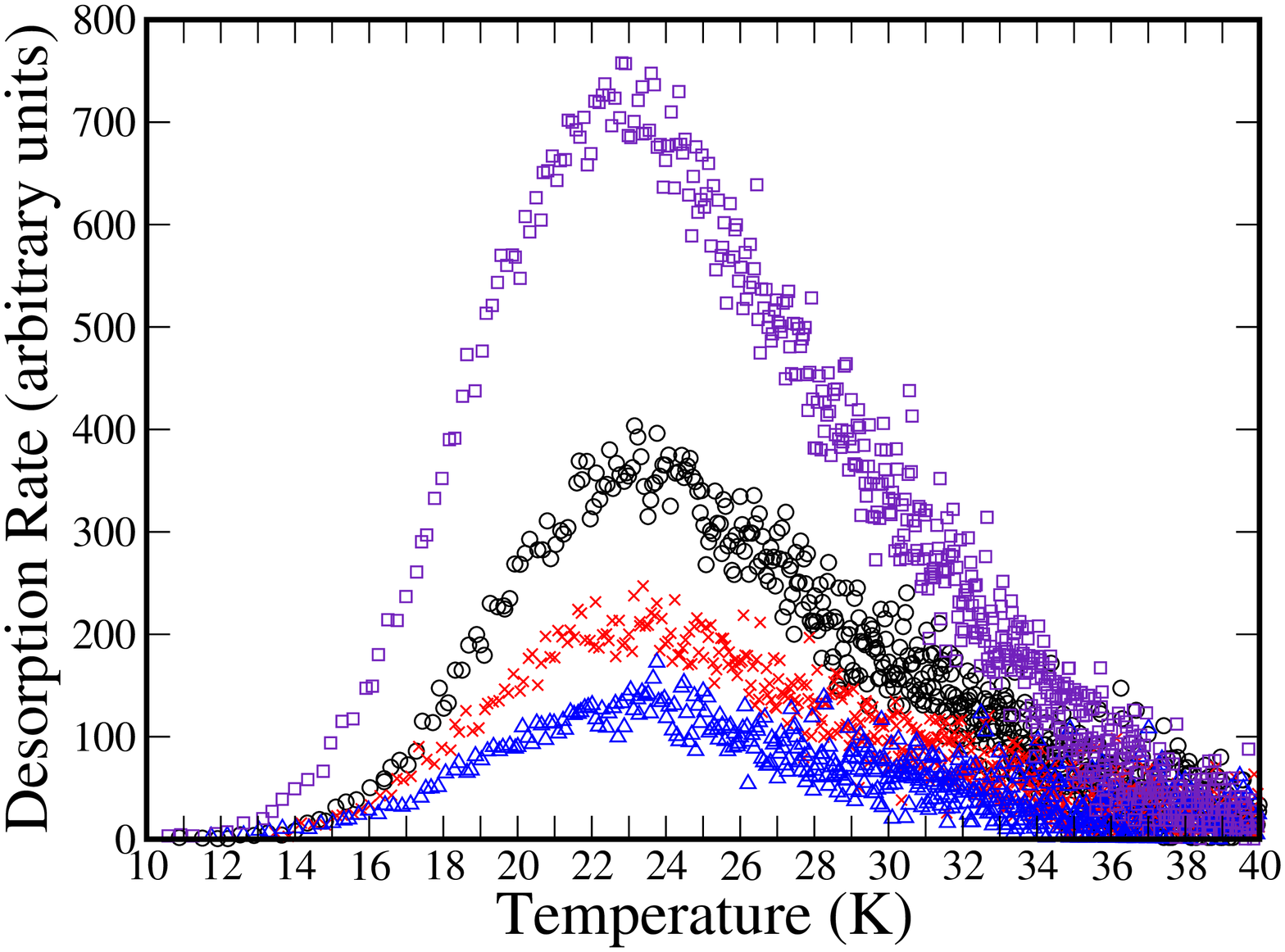}
\caption{
TPD traces obtained after
irradiation with H+D atoms at a sample temperature
of 10K. The irradiation times are
30 (triangles), 60 ($\times$), 120 (circles) and 240 (squares) seconds.
}
\label{fig:4}
\end{figure}

In Fig. 
\ref{fig:1}
we show a typical TPD trace 
of HD desorption, taken 
after irradiating the sample, while at 5.6K,
with HD molecules ($\times$)  
and 
with H and D atoms (circles). 
Both curves exhibit a strong 
peak at about 15-16K and a lower but broad peak
(or a shoulder) around 21K.  
The broad peaks reflect the fact that the surface 
is disordered. For comparison, a TPD trace taken 
from a polycrystalline 
olivine sample is much narrower and has a maximum 
at a lower temperature
\cite{Perets2007}.

In Fig.  
\ref{fig:2}
we present the desorption rates of HD
after irradiation with 
HD molecules ($\times$) 
and 
H+D atoms (circles)
at a sample temperature of $\sim 10$K. 
These TPD traces exhibit two peaks.
The main peak coincides with the high temperature peak
of Fig. 
\ref{fig:1}. 
In addition, there is a smaller peak or a
shoulder at $\sim$31-32K, 
which does not appear in Fig. 
\ref{fig:1}. 
However, it turns out that the TPD curve of Fig. 1
was recorded only up to 24K. 
Therefore, we cannot reject the possibility that 
a third peak, around 31-32K also appears
under the conditions of Fig. 1.

The experiments with HD irradiation provide direct
information on the energy barriers for the 
desorption of HD molecules. 
Combining the two sets of experiments,
we find that there are three types of adsorption 
sites for molecules: shallow, medium and deep.
When HD molecules are irradiated at 5.6K they 
reside in the shallow and medium depth sites;
when the molecules are irradiated at 10K they
mostly reside reside in the medium and deep sites.
This indicates that during irradiation at 10K the HD molecules have
sufficient mobility to hop from the shallow sites
to deeper sites before they desorb.
The similarity between the TPD curves obtained with
HD and H+D irradiations indicates
that molecules just formed on the surface 
quickly thermalize with the surface temperature
and occupy the 
same adsorption sites
as those molecules  
deposited from the gas phase 
\cite{Perets2007}. 

In Fig. 
\ref{fig:3}
we present several 
TPD traces obtained after
irradiation with HD molecules at a sample temperature
of 10K. The irradiation times are
30 (triangles), 60 ($\times$), 120 (circles) and 240 (squares) s.
In all cases the main peak is at 22K, as expected
for irradiation with molecules, leading to first order kinetics.
The high temperature peak appears around 31-32K 
for all exposures.

In Fig. 
\ref{fig:4}
we present several TPD traces obtained after
irradiation with H+D atoms at a sample temperature
of 10K. The irradiation times are
30 (triangles), 60 ($\times$), 120 (circles) and 240 (squares) s.
The location of the peak is clearly the same for all
exposures, indicating a first order kinetics. 
This peak temperature (around 23K) is only slighly higher 
than the 22K peak obtained for HD irradiation.
However, the peak obtained for H+D irradiation 
exhibits a broader high temperature wing, which
merges with the second peak.
The first order feature of the H+D peak
for irradiation at of 10K
may indicate that some of the molecules are
formed during irradiation or at early stages
of the TPD run and quickly equilibrate with
the surface.
However, the broad high temperature wing may
indicate that some other molecules are formed
at later stages of the TPD run.
In this case, the high temperature peak
(which is difficult to resolve)
may exhibit a combination of first and second order features.
The experimental data was analyzed using the rate
equation models described below.
The results for the energy barriers for diffusion
and desorption are summarized in Table I.

\section{The Rate Equation Model}
\label{sec:model}

Consider an experiment in which beams of H and D atoms
are irradiated on a surface.
Atoms that stick to the
surface hop as random walkers.
The hopping atoms may either encounter each other and form 
H$_2$, HD and D$_2$
molecules, or desorb from the surface in atomic form. 
As the sample temperature is raised, both the hopping
and desorption rates quickly increase.
The fluxes of H and D 
are denoted by
$f_{\rm H}$
and 
$f_{\rm D}$,
respectively
[in units of monolayer (ML) $s^{-1}$].
These fluxes are related to the beam intensities
by
$f_{\rm H} = F_{\rm H}/s$
and
$f_{\rm D} = F_{\rm D}/s$,
where $s$ (cm$^{-2}$)
is the density of adsorption sites
on the surface.

The process of molecular hydrogen formation is described by
a rate equation model, which includes 
three surface reactions, namely
${\rm H}+{\rm H} \rightarrow {\rm H}_2$, 
${\rm H}+{\rm D} \rightarrow {\rm HD}$ 
and
${\rm D}+{\rm D} \rightarrow {\rm D}_2$. 
The surface coverages of adsorbed H and D atoms are denoted by
$n_{\rm H}$ (ML)
and
$n_{\rm D}$ (ML),
respectively.
Similarly, 
the coverages of the adsorbed 
H$_2$, D$_2$ and HD molecules
are denoted by
$n_{\rm H_2}$, $n_{\rm D_2}$ and $n_{\rm HD}$ (ML).
The time derivatives of the coverages of adsorbed atoms 
are given by the rate equations 

\begin{eqnarray}
\frac{dn_{\rm H}}{dt}&=&
  f_{\rm H}
- W_{\rm H} n_{\rm H} 
- 2 a_{\rm H} n_{\rm H}^2
- (a_{\rm H}+a_{\rm D}) n_{\rm H} n_{\rm D} 
\nonumber \\
\frac{dn_{\rm D}}{dt}&=&
  f_{\rm D}
- W_{\rm D} n_{\rm D} 
- 2 a_{\rm D} n_{\rm D}^2
- (a_H+a_D) n_{\rm H} n_{\rm D}. 
\label{eqn:rateqHD}
\end{eqnarray}

\noindent
The time derivatives of the
coverages of adsorbed molecules are given by

\begin{eqnarray}
\frac{dn_{\rm H_2}}{dt} &=& 
  a_{\rm H} n_{\rm H}^2
- W_{\rm H_2} n_{\rm H_2} 
\nonumber \\
\frac{dn_{\rm D_2}}{dt} &=& 
  a_{\rm D} n_{\rm D}^2 
- W_{\rm D_2} n_{\rm D_2} 
\nonumber \\
\frac{dn_{\rm HD}}{dt}  &=& 
  (a_{\rm H}+a_{\rm D}) n_{\rm H} n_{\rm D} 
- W_{\rm HD} n_{\rm HD}. 
\label{eqn:rateqHD2}
\end{eqnarray}

\noindent
The first terms on the right hand sides of
Eqs.~(\ref{eqn:rateqHD}) 
represent the incoming
flux.
For simplicity, we ignore the Langmuir rejection.
The second terms in 
Eqs.~(\ref{eqn:rateqHD}) 
represent the desorption of H and D atoms from the
surface,
while the second terms in 
Eqs.~(\ref{eqn:rateqHD2})
represent the desorption of molecules.
The desorption coefficients are 

\begin{equation}
W_{\rm X}=  \nu \exp (- E_{\rm X}^{\rm des} / k_{\rm B} T)  
\end{equation}

\noindent
where $\nu$ is the attempt rate 
(standardly taken to be $10^{12}$ $s^{-1}$), 
$E_{\rm X}^{\rm des}$ 
is the energy barrier for desorption 
of species
$X$, 
where X$=$ H, D, H$_2$, HD or D$_2$, 
and $T$ (K) is the surface temperature.
The third and fourth terms in 
Eqs.~(\ref{eqn:rateqHD})
and the first terms in 
Eqs.~(\ref{eqn:rateqHD2}) 
account for the formation of H$_2$, HD and D$_2$
molecules,
where

\begin{equation}
a_X =  \nu \exp (- E_{\rm X}^{\rm diff} / k_{\rm B} T) 
\end{equation}

\noindent
is the hopping rate of 
atoms of species 
X $=$ H, D
between adsorption sites 
on the surface
and
$E_{\rm X}^{\rm diff}$ 
is the energy barrier for hopping.
Here we assume that there is no energy barrier for the
formation of hydrogen molecules on the surface,
namely two hydrogen atoms that encounter each other
form a molecule. 
The desorption rate 
$R_{\rm X}$
(ML s$^{-1}$)
of molecular species
X is $R_{\rm X} = W_{\rm X} n_{\rm X}$.
In particular, 
the rate of HD desorption is given by

\begin{equation}  
R_{\rm HD} =  W_{\rm HD} n_{\rm HD}.
\label{eq:HDprod}
\end{equation}

To simplify the analysis and reduce the number of fitting parameters
it is desireable to use a model which includes only one isotope
of hydrogen. 
Here we examine the effect of the isotopic difference
on the TPD results.
Due to its higher mass, the zero point energy for
adsorbed D atoms is lower than for H atoms.
As a result, the energy barriers for diffusion and
desorption of D atoms are expected to be higher than
for H atoms.
The evaluation of the difference in the 
activation energies 
between the H and D isotopes
requires detailed knowledge of 
the atom-surface potentials. Such potentials are not
available for the amorphous surfaces of interest here. 
However, one can obtain a rough idea about the isotope
effect by considering the difference between
the activation energies of H and D atoms
on a graphite surface. 
In this case, measurements show that
the laterally averaged binding energy 
on the basal plane of graphite,
in the ground state level
is 31.6 meV for H atoms
and it is 3.8 meV higher for D atoms
\cite{Ghio1980}.
The isotopic difference in the energy barriers
for diffusion
is expected to be at most as large as 
the difference in the energy barriers for
desorption.

\begin{figure} 
\includegraphics[scale=0.55,angle=0]{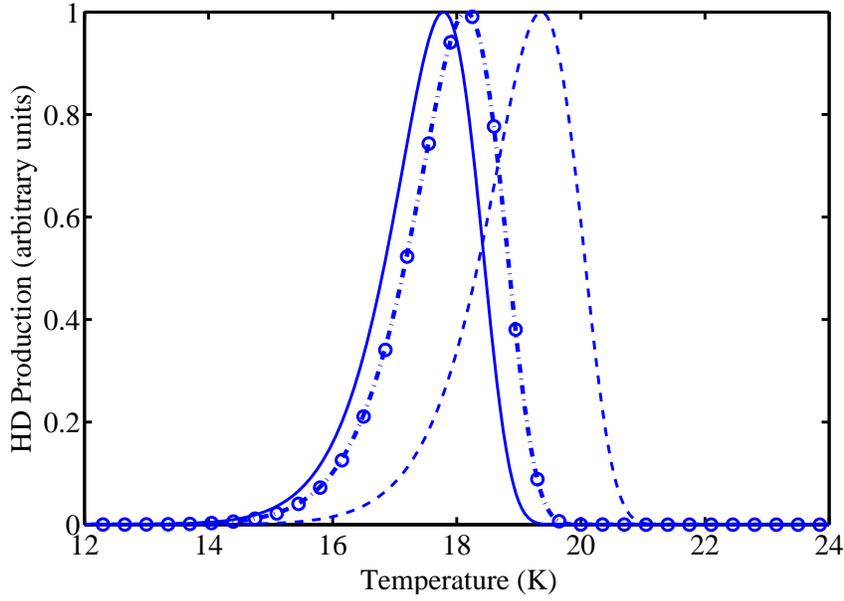}
\caption{
Simulated TPD curves of HD production and desorption
obtained from Eqs. (\ref{eqn:rateqHD}),
following irradiation by H+D atoms
at 5.6K and heating at a constant rate of 0.5K/s. 
In this Figure it is assumed that HD molecules desorb 
upon formation.
Four choices of parameters are shown:
the parameters of H and
D are identical, and are given by Table I
(solid line);
the desorption energy barrier of D is raised by 4.0 meV
(dashed-dotted line);
both the desorption and diffusion energy barriers of D
are raised by 4.0 meV (circles);
the desorption and diffusion energy barriers of both H and
D are raised by 4.0 meV (dashed line).
The results indicate that
the higher binding energy of D atoms has little
effect on the TPD curves.
The peak heights are normalized to 1, to make it
easier to compare the peak locations.
}
\label{fig:5}
\end{figure}

Here we examine the effect of increasing the energy barriers
for D diffusion and desorption, keeping those of H atoms 
unchanged. 
In Fig. 
\ref{fig:5} 
we present the TPD curves obtained 
from the rate equation model,
following irradiation at 5.6K and heating
at a constant rate of 0.5K/s. 
Four curves are shown:
the TPD results obtained when the parameters of H and
D are identical, and given by Table I
(solid line);
the energy barrier for desorption of D 
atoms is raised by 4.0 meV
(dashed-dotted line);
the energy barriers for
desorption and diffusion of D atoms
are raised by 4.0 meV (circles);
and
the energy barriers for desorption and diffusion 
of both H and D 
are raised by 4.0 meV (dashed line).
We find that raising the energy barriers 
for desorption (and diffusion) 
of D atoms gives rise to only a 
slight shift of the peak to a higher temperature.
We conclude that the TPD curves are mainly determined
by the isotope that interacts more weakly with the surface.
It is thus appropriate to use a simplified model
which includes only one isotope.
The energy barriers obtained should be interpreted as the
barriers for H atoms.

In the models we assume a given density of adsorption sites
on the surface. 
In terms of the adsorption of H atoms, 
all the adsorption sites are assumed to be identical,
where the energy barrier for H diffusion is
$E_{\rm H}^{\rm diff}$
and the barrier for desorption
is
$E_{\rm H}^{\rm des}$.
This assumption is justified by the fact that 
such a simple model provides good fits to the
TPD curves obtained after H+D irradiation.
If there is a distribution of diffusion energy barriers,
the hydrogen atoms are expected to 
be trapped most of the time in the deep adsorption sites.
Thus, the diffusion-induced recombination is expected
to be dominated by the deepest wells.
This means that the parameters obtained from fitting
the TPD curves to the rate equation model characterize
the upper edge of the distribution of energy barriers.
A more complete model that provides a connection between
the roughness and the distribution of energy barriers
for H diffusion and desorption
was studied in Ref.
\cite{Cuppen2005}.

As for the adsorption of hydrogen molecules, we assume that
the adsorption sites may differ from each other.
The energy barriers for desorption 
of HD molecules can be obtained
directly from the TPD curves that follow HD irradiation.
These curves, obtained for different
surface temperatures during irradiation,
reveal a total of three peaks, indicating that
there are three types of adsorption sites.  
In general, we assume that the adsorption
sites are divided into
$J$ types
according to the binding energies of trapped HD molecules.
Ignoring the isotopic differences for molecules,
the energy barrier for desorption of H$_2$ molecules
from an adsorption site of type $j$ is
$E_{\rm H_2}^{\rm des}(j)$, where
$j=1,\dots,J$. 
The parameter
$\mu_j$ 
represents the fraction of
the molecules that are trapped
in sites of type $j$
upon formation or adsorption
and
$\sum_j \mu_j = 1$.
Let $n_{\rm H_2}(j)$ (ML) be
the coverage of H$_{2}$ molecules that are 
trapped in adsorption sites of type $j$, where 
$j=1,\dots,J$.
The rate equation model takes the form
\cite{Perets2005}

\begin{subeqnarray}
\label{eq:N2} 
& \dot{n}_{{\rm H}} & =  
f_{\rm H}\left(1 - n_{\rm H} \right) 
-W_{\rm H}n_{\rm H} - 2a_{\rm H} {n_{\rm H}}^{2}
\slabel{eq:NH} \\
& \dot{n}_{\rm H_2}(j) & 
= \mu_{j} a_{\rm H} {n_{\rm H}}^{2} - W_{\rm H_2}(j)n_{\rm H_2}(j),
\slabel{eq:N2(j)} 
\end{subeqnarray}

\noindent
where $j=1,\dots,J$.
The first term on the right hand side of
Eq.~(\ref{eq:NH})
represents the incoming
flux in the Langmuir kinetics.
In this scheme H atoms deposited on top of H atoms
already on the surface are rejected.
There are indications that adsorbed H$_2$ molecules
do not lead to Langmuir rejection
of H atoms
\cite{Govers1980}.
The parameter
$f_{\rm H}$ represents an
{\em effective} flux (in units of ML s$^{-1}$),
namely it
already includes the possibility of a temperature
dependent sticking coefficient.
The second term in
Eq.~(\ref{eq:NH})
represents the desorption of H atoms from the
surface.
The third term in
Eq.~(\ref{eq:NH})
accounts for the depletion of the H atoms 
on the surface due to diffusion-mediated recombination 
into H$_{2}$ molecules.
Eq. (\ref{eq:N2(j)})
accounts for
the population of molecules on the surface.
The first term on the right hand side 
represents the formation of H$_2$ molecules
that become adsorbed in a site of type 
$j$.
The second term in
Eq. (\ref{eq:N2(j)})
describes the desorption of H$_2$ molecules from
sites of type $j$,
where

\begin{equation}
W_{\rm H_2}(j)  =  
\nu \exp [- E_{\rm H_2}^{\rm{des}}(j) / k_{B} T]
\label{eq:Wk}
\end{equation}

\noindent
is the H$_2$ desorption coefficient.
The H$_2$ production rate $R_{\rm H_2}$ 
(ML s$^{-1}$)
is given by:

\begin{equation}
R_{\rm H_2}  =  \sum_{j=1}^{J}{W_{\rm H_2}(j) n_{\rm H_2}(j)}.
\label{eq:Production}
\end{equation}

\noindent
To analyze the diffusion of H$_2$ molecules,
one needs to use a more complete model
\cite{Perets2005}.
This model includes additional parameters,
namely the energy barriers
$E_{\rm H_2}^{\rm diff}(j)$
for hopping of H$_2$ molecules out of 
sites of type $j$,
as well as the partial densities 
$s_j$ (cm$^{-2}$)
of such sites.
It is difficult to extract, 
from the experimental data,
unique values for all these parameters with sufficient 
confidence.

\section{Analysis of the Experimental Results}
\label{sec:analysis}

In Ref.
\cite{Perets2007} we presented 
a series of TPD traces obtained after
irradiation of H and D atoms on an amorphous
silicate sample 
at low surface temperatures (5.6K).
Each trace exhibits a large peak at a lower temperature
and a small peak at a higher temperature.
The location of the low temperature peak shifts to the 
right as the irradiation time is reduced, 
suggesting second order kinetics.
We should note that a somewhat similar behavior
was found in TPD experiments in which
D$_2$ molecules were irradiated on amorphous water ice
\cite{Dulieu2005,Amiaud2006}.
In that case the second-order like behavior was interpreted
as a result of the saturation of the deepest adsorption sites.
However, in our experiments the coverage of hydrogen atoms
on the surface is very low. 
Therefore, it is unlikely that saturation effects play
a significant role.

This observation indicates that the molecules are formed during
the heating and not during the irradiation stage.
Thus, the adsorbed hydrogen atoms are
immobile during irradiation at 5.6K.
This means that tunneling alone is not sufficient in order
to provide significant mobility to the adsorbed hydrogen atoms,
and their mobility is dominated by thermal activation.
A similar conclusion was reached from analysis of data 
from polycrystalline olivine
\cite{Katz1999}.

Assuming that the 
Langmuir rejection mechanism applies,
hydrogen atoms that are deposited on top of already 
adsorbed atoms are rejected.
This provides a prediction
for the coverage of adsorbed atoms after 
irradiation time $t_0$.
Taking the Langmuir rejection into account, the
coverage is given by
\cite{Biham2001}

\begin{equation}
n_{\rm H}(t_0) = 1 - \exp(-f_{\rm H} \cdot t_0).
\label{eq:Lrejection}
\end{equation}

\noindent
In the experiment, the coverage after irradiation
can be evaluated using the total yield of
HD molecules in each TPD run.
The total yields for irradiation at 5.6K and several
exposure times were evaluated and fitted 
according to
Eq.
(\ref{eq:Lrejection}).
It was found 
that the flux of incoming atoms is
$f_{\rm H} = 7.0 \times 10^{-4}$ 
(in ML s$^{-1}$). 
Since the beam intensities $F_{\rm H}$ and $F_{\rm D}$  are known, 
one can use the relations $f_X = F_X/s$,
where X=H, D,
to obtain the density of adsorption sites. 
It was found that the density of
adsorption sites on the amorphous silicate sample
is
$s=7 \times 10^{14}$ 
(sites cm$^{-2}$). 

The experimental results were fitted using
the rate equation model described above. 
The parameters for the diffusion
and desorption of hydrogen atoms and molecules 
on the amorphous silicate surface were obtained. 
These include the energy barrier 
$E_{\rm H}^{\rm diff}=35$ (meV) 
for the diffusion of H atoms
and the barrier
$E_{\rm H}^{\rm des}=44$ (meV)  
for their desorption.
The value obtained for the energy barrier
for desorption
should be considered only as a lower bound, 
because the TPD results are insensitive 
to variations in 
$E_{\rm H}^{\rm des}$,  
as long as it is higher than the reported value.
The desorption energy barriers of HD molecules adsorbed
in shallow (lower temperature peak)
and deep (higher temperature peak) 
sites, are given by 
$E_{\rm H_{2}}^{\rm des}(1)=35$ 
and
$E_{\rm H_{2}}^{\rm des}(2)=53$ (meV). 

The rate equation model is integrated using a Runge Kutta stepper. 
For any given choice of the parameters, one obtains a set of TPD
curves for the different irradiation times used in the experiments.
The actual time dependence of the temperature, T(t), 
recorded during the experiment,
is taken into account as follows. 
The rate equations [Eqs. (6)] 
are integrated numerically using a Runge Kutta stepper. 
The rate constants for diffusion and desorption 
[given by Eqs. (3), (4) and (7)], which depend on the
temperature, are adjusted during the integration according to
the experimentally recorded temperature curve, T(t). 
Thus the simulation fully reflects the physical conditions
during the experiment.

In the first step, the barriers 
$E_{{\rm H_{2}}}^{{\rm des}}(j)$, $j=1,\dots,J$, 
for the desorption of molecules
are obtained using the results of the experiments in which
HD molecules are irradiated on the surface.
To obtain better fits of the peak shape, 
suitable Gaussian distributions of energy barriers 
around these two values
can be incorporated 
(see Figs. 1 and 2 in Ref. 
\cite{Perets2007}).
However, even with these Gaussian distributions it is
difficult to fit the leading edge of the TPD curves.
This may indicate that the distribution of the 
energy barriers is non-symmetric and exhibits a broader
tail on the lower side.

A wide range of energies can be expected for 
a morphologically heterogeneous surface. 
For example, in a calculation of H adsorption 
energies and diffusion barriers on a water 
cluster simulating the surface of amorphous water ice, 
Buch and co-workers 
obtained Gaussian distribution for these energies
\cite{Buch1991a,Buch1991b}.

Using a single value of each barrier, one obtains
fits which consist of rather narrow peaks.
Their locations coincide with
the experimental peaks but they do not capture the tails.

The results of the H+D experiments (circles in Fig. 1)
are fitted using the rate equation model (solid line)
with the activation energies specified above.
The weights of the molecular 
adsorption sites are chosen to be
$\mu_1=0.6$
and
$\mu_2=0.4$.
These values reflect the total relative yields
of desorbed HD molecules in the two peaks.
The results of H+D irradiation at 10K,
shown in Fig.
\ref{fig:2},
were also fitted using
the rate equation model.
The activation energies,
$E_{\rm H}^{\rm diff}$
and 
$E_{\rm H}^{\rm des}$
are the same as above.
However, in this case it is found that the 
molecular adsorption sites
indexed by $j=1$ are too shallow to trap molecular
hydrogen when the irradiation is done at 10K.
To account for this fact we impose
$\mu_1=0$.
The TPD curve is fitted with
$E_{\rm H_2}^{\rm des}(2)=53$ 
(as above)
and
$E_{\rm H_2}^{\rm des}(3)=75$ (meV).
The weights are
$\mu_2=0.5$
and
$\mu_3=0.5$.

\section{Discussion and Applications}
\label{sec:applications}

The processes of molecular hydrogen formation 
on amorphous silicate surfaces at low temperatures 
are based on physisorption forces. 
The activation energies that we obtained show that 
these values are higher than expected for other 
physisorption interactions that have been measured 
between H and the surface of single crystals
\cite{Vidali1991}.
This is probably due to the more complex chemical and morphological 
composition of the amorphous silicate samples.
Using the parameters obtained from the experiments we now calculate
the recombination efficiency of hydrogen on 
amorphous silicate surfaces
under interstellar conditions. 
The recombination efficiency is defined as
the fraction of the adsorbed hydrogen atoms 
which come out as molecules, namely 
\cite{Biham1998}

\begin{equation}
\eta = \frac{R_{\rm H_2}}{(f_{\rm H}/2)}.
\label{eq:eta}
\end{equation}

\noindent
Under steady state conditions, the recombination rate
$R_{\rm H_2}$ (ML s$^{-1}$)
is given by 
$R_{\rm H_2} = a_{\rm H} {n_{\rm H}}^2$,
where

\begin{equation}
n_{\rm H} = \frac{1}{4 a_{\rm H}} 
\left[-(W_{\rm H} + f_{\rm H}) 
+ \sqrt{(W_{\rm H} + f_{\rm H})^2 + 8a_{\rm H} f_{\rm H}} \right].
\end{equation}

\noindent
Within the assumptions of this model, the binding 
energies 
$E_{\rm H_2}^{\rm des}$
of H$_2$ molecules to the surface do not affect the rate of 
H$_2$ formation. They only determine the residence time of 
H$_2$ molecules on the grain and thus affect the
steady state coverage of H$_2$ molecules on the surface.

\begin{figure} 
{\large (a)}
\includegraphics[scale=0.45,angle=0]{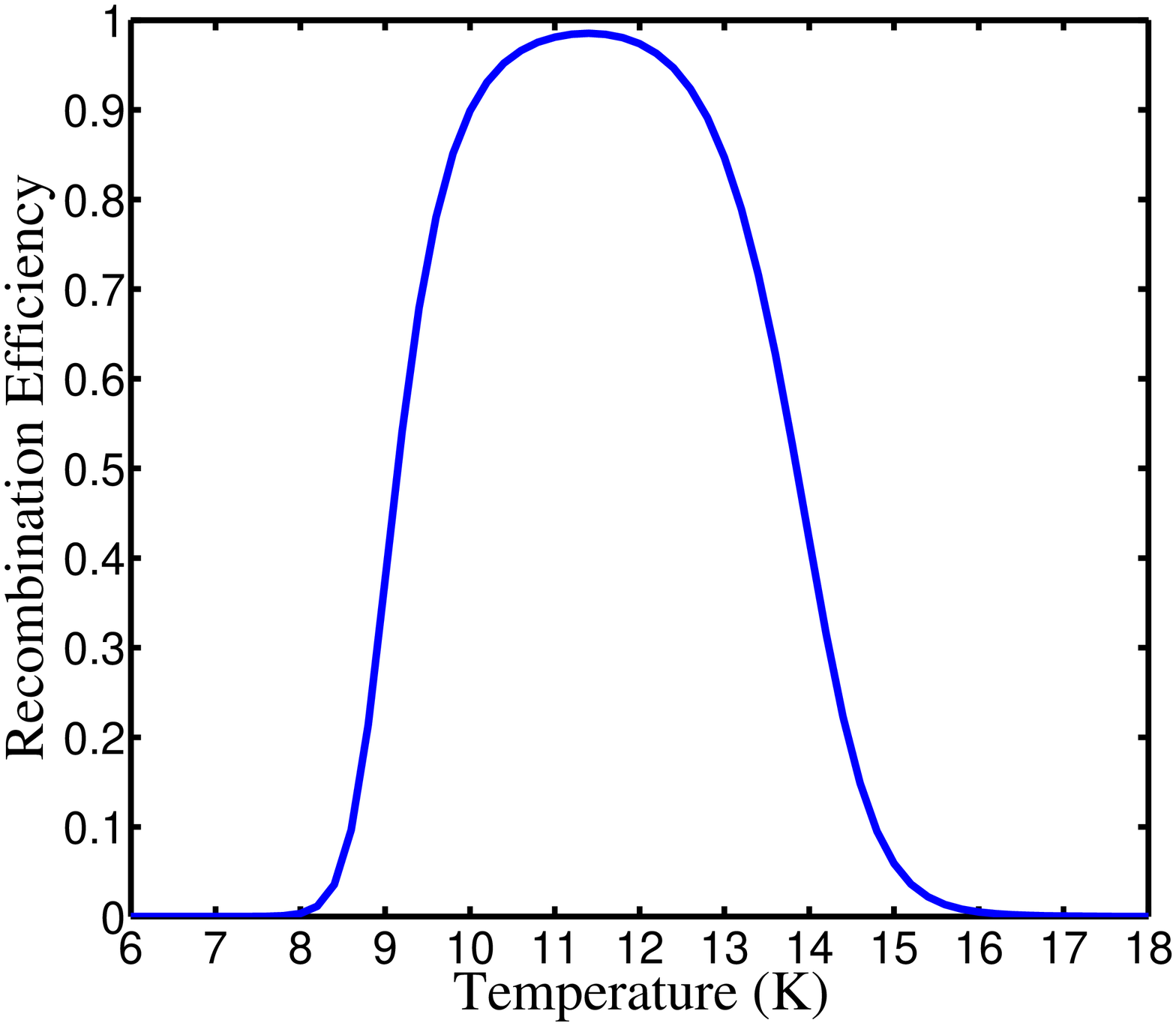} \\
{\large (b)}
\includegraphics[scale=0.45,angle=0]{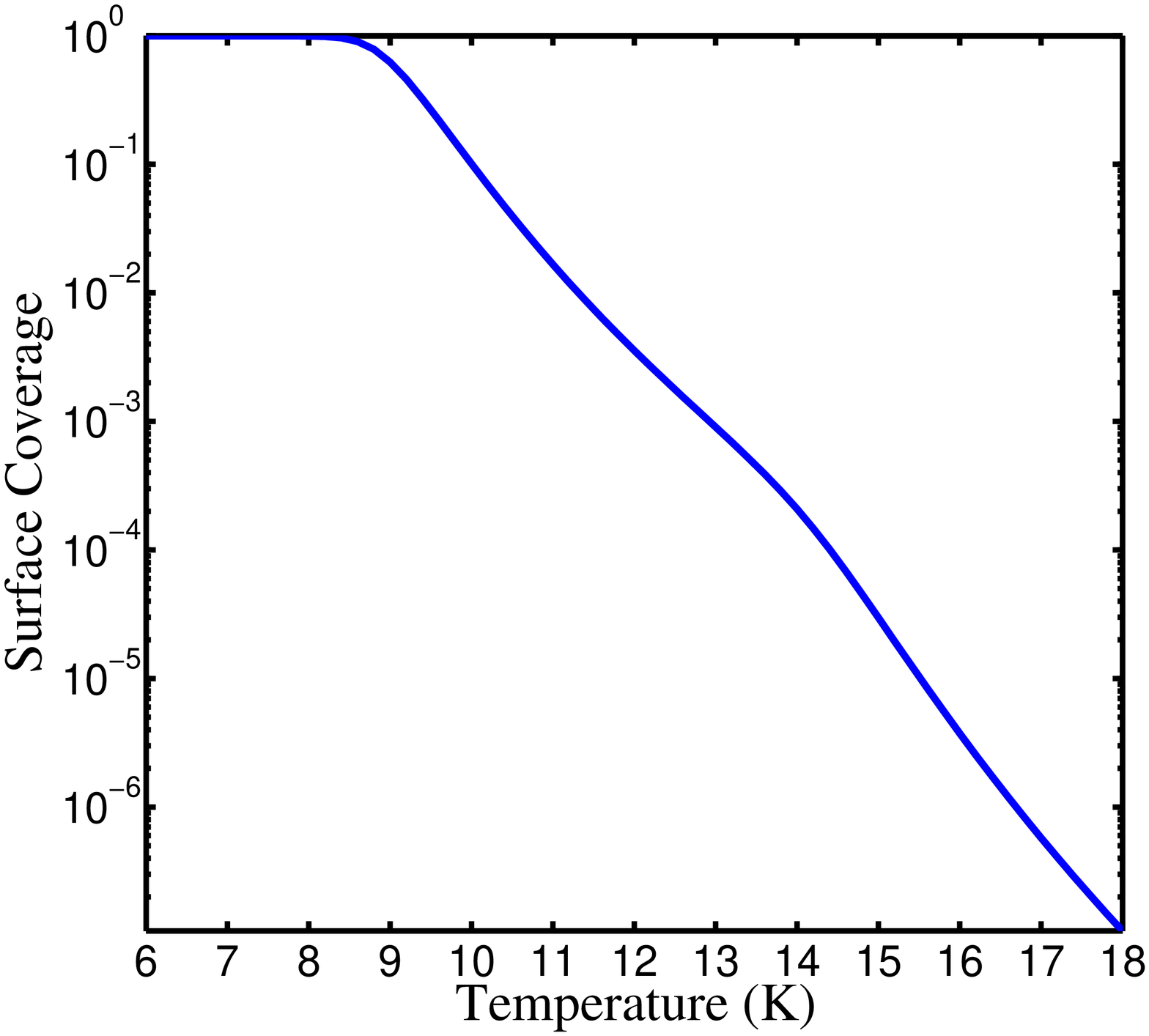}
\caption{
(a) Calculated recombination efficiency 
and surface coverage (b) of
hydrogen on the amorphous silicate sample vs. surface temperature,
under steady-state conditions. 
A temperature window of high efficiency is found.
Within this window, the surface coverage is reduced
from a full monolayer to about 10$^{-4}$ ML.
}
\label{fig:6}
\end{figure}

In Figs.  
\ref{fig:6}(a)
and
\ref{fig:6}(b)
we present the 
recombination efficiency
and the coverage of H atoms, respectively,
vs. surface temperature for 
an amorphous silicate sample under flux of
$f_{\rm H}=$ 5.2 $\times$ $10^{-9}$ (ML s$^{-1}$). 
This flux is within the typical
range for diffuse interstellar clouds, 
where bare amorphous silicate grains are expected to play a crucial role
in H$_2$ formation. 
This flux
corresponds to gas density of 100 
(atoms cm$^{-3}$), 
gas temperature of 100K and 
a density of
$7 \times 10^{14}$ adsorption
sites per cm$^{2}$ on the grain surface.
A window of high recombination efficiency is found between 9-14K,
compared to 6-10K for polycrystalline silicate under similar conditions. 
The surface coverage of H atoms decreases dramatically within
the efficiency window, from nearly full monolayer at 9K
to about 10$^{-4}$ ML at 14K.
This is due to the fact that as the temperature increases,
the surface mobility of H atoms is enhanced and their
residence time before recombination is reduced.

In general, the high efficiency window
for hydrogen recombination 
is bounded from below by
\cite{Biham2002}

\begin{equation}
T_{\rm low}(f_{\rm H}) = 
\frac{E_{\rm H}^{\rm diff}}{k_{\rm B}(\ln \nu - \ln f_{\rm H})},
\end{equation}

\noindent
and from above by

\begin{equation}
T_{\rm high}(f_{\rm H}) = 
\frac{{2 E_{\rm H}^{\rm des}-E_{\rm H}^{\rm diff}}}
{k_{\rm B} (\ln \nu - \ln f_{\rm H})} . 
\label{eq:gvul12}
\end{equation}

\noindent
These bounds are obtained by solving
Eq. (\ref{eq:NH})
under steady state conditions
for recombination efficiency of 50\%.
Thus, inside the window the efficiency is
higher than 50\%, while outside it is lower 
than 50\%.
The recombination efficiency declines sharply
near these bounds.
The width of the high efficiency window is
proportional to the difference,
$E_{\rm H}^{\rm des} - E_{\rm H}^{\rm diff}$,
between the diffusion and desorption energy barriers
of H atoms.
The location of this window exhibits a logarithmic
dependence on the flux $f_{\rm H}$ and
slowly shifts to higher temperatures as $f_{\rm H}$ 
increases.

At temperatures higher than 
$T_{\rm high}$
atoms desorb from the surface before they have sufficient time to
encounter each other. 
At temperatures 
lower than
$T_{\rm low}$
diffusion is suppressed and
the surface is saturated by immobile H atoms.
As a result,
the Langmuir-Hinshelwood mechanism is no longer efficient.
Saturation of the surface with immobile
H atoms 
might render the Eley-Rideal mechanism more efficient
in producing some 
recombination 
\citep{Katz1999,Perets2005}. 
Our results
thus indicate that the recombination efficiency of hydrogen on 
amorphous silicates is high
in a temperature range, which is relevant to interstellar clouds.
Therefore, amorphous silicates 
seem to be good candidates for interstellar
grain components on which hydrogen recombines 
with high efficiency.

To apply the results of the experiments to H$_2$ formation
in the ISM, consider for simplicity a spherical grain
of radius $r$.
The cross section of such grain is
$\sigma = \pi r^2$
and the number of adsorption sites on its surface is
$S=4 \pi r^2 s$.
We denote by $N_{\rm H}$
the number of H atoms adsorbed on such grain.
Its time dependence is given by

\begin{equation}
\dot N_{\rm H} = 
  F_{\rm H} \sigma 
- W_{\rm H} N_{\rm H}
- A_{\rm H} N_{\rm H}^2.
\label{eq:rateN}
\end{equation}

\noindent
The incoming flux density is given by
$F_{\rm H} = n_{\rm H}^{\rm gas} v_{\rm H}$,
where 
$n_{\rm H}^{\rm gas}$ (cm$^{-3}$) 
is the density of H atoms in the gas phase
and
$v_{\rm H}$
is their average velocity.
The parameter
$A_{\rm H} = a_{\rm H} / S$, 
is the sweeping rate,
which is approximately the inverse of the time it takes  
for an H atom to visit nearly all  
the adsorption sites on the grain surface
(for a more precise evaluation of the sweeping rate see Refs. 
\cite{Krug2003,Lohmar2006}). 

The production rate of H$_{2}$ molecules on a single grain,
is given by  

\begin{equation}
R_{\rm H_2}^{\rm gr} =  A_{\rm H}  {N_{\rm H}}^2. 
\label{eq:R_H2}
\end{equation}

\noindent
Under given flux and surface temperature, 
for grains that are large enough to hold many H atoms,
Eqs. (\ref{eq:N2})
provide a good description of the recombination process.
However, in the limit of small grains and low flux,
$N_{\rm H}$ 
may be reduced to order unity or less.
Under these conditions,
Eqs. (\ref{eq:N2})
become unsuitable, 
because they ignore the discrete nature of the population
of adsorbed atoms and its fluctuations.

To account for the reaction rates on small grains,
a modified set of rate equations
was introduced 
and applied  
to a variety of chemical
reactions
\cite{Caselli1998,Shalabiea1998}.  
In these equations the rate coefficients are modified in a semi-empirical
way taking into account the effect of the finite grain 
size on the recombination process.
The modified rate equations take into account correctly
the length scales involved in the recombination process
on small grains
\cite{Biham2002}.
However, they still 
involve only the average 
values and ignore fluctuations, thus providing only
approximate results for the reaction rates on small grains.

To account correctly for the 
effects of fluctuations on the
recombination rate on small grains,
simulations using the master equation are required.
The dynamical variables
of the master equation are the probabilities
$P(N_{\rm H})$
of having a population of $N_{\rm H}$ hydrogen atoms 
on the grain.
In the case of hydrogen recombination, 
the master equation takes the form
\cite{Biham2001,Green2001}

\begin{eqnarray}
\dot {P} (N_{\rm H} ) &=& F_{\rm H}  
\left[ P (N_{\rm H} -1) - P (N_{\rm H} ) \right]
+ W_{\rm H}  
\left[ (N_{\rm H} +1) P(N_{\rm H} +1) - N_{\rm H}  P(N_{\rm H} ) \right] 
\nonumber \\
&+& A_{\rm H}  [ (N_{\rm H} +2) (N_{\rm H} +1) P(N_{\rm H} +2) 
- N_{\rm H} (N_{\rm H} -1) P (N_{\rm H} ) ],
\label{eq:master_1specie}
\end{eqnarray}

\noindent
where $N_{\rm H}=0,1,2,\dots,N_{\rm H}^{\rm max}$
(the master equation must be truncated in order 
to keep the number of equations finite).
The first term on the right hand side of 
Eq. (\ref{eq:master_1specie})
describes the effect of the incoming flux. 
The probability $P(N_{\rm H} )$ increases when an H atom is adsorbed by
a grain that already has $N_{\rm H} -1$ adsorbed H atoms, and decreases when 
it is adsorbed on a grain with $N_{\rm H}$ atoms.
The second term accounts for the desorption process. 
The third term describes the recombination process. 
The recombination rate is proportional to the number of pairs of
H atoms on the grain, namely $N_{\rm H} (N_{\rm H} -1) / 2$.
Therefore, the H$_2$ production rate per grain
can be expressed in terms
of the first two
moments of $P(N_{\rm H})$, 
according to

\begin{equation}
R_{\rm H_2}^{\rm gr} = A_{\rm H} 
( \langle N_{\rm H}^2 \rangle -  \langle N_{\rm H} \rangle ),
\label{eq:master_production}
\end{equation}

\noindent
where

\begin{equation}
\langle N_{\rm H}^k \rangle = 
\sum_{N_{\rm H}=0}^{N_{\rm H}^{\rm max}}
{N_{\rm H}}^k P(N_{\rm H}).
\label{eq:momentk}
\end{equation}

The master equation can be simulated either by direct
numerical integration or by a stochastic implementation
via Monte Carlo (MC) methods 
\cite{Gillespie1976,Charnly2001}.
A significant advantage of the
direct integration over the MC approach
is that the equations
can be easily coupled to the  
rate equations of gas-phase chemistery. 
However, the number of coupled equations increases exponentially 
with the number of reactive species, 
making direct integration infeasible for complex reaction
networks of multiple species
\cite{Stantcheva2002,Stantcheva2003}.
To solve this problem, two approaches have been proposed:
the multiplane method 
\cite{Lipshtat2004,Barzel2007c}
and the moment equations
\cite{Lipshtat2003,Barzel2007a,Barzel2007b}.

The multiplane method 
\cite{Lipshtat2004}
is based on breaking the network into a set of 
maximal fully connected subnetworks (maximal cliques).
It involves an approximation, 
in which the correlations between 
pairs of species that react with each other are maintained, while
the correlations between non-reacting pairs are neglected
\cite{Barzel2007c}. 
The result is a set of lower dimensional master equations,
one for each clique, with suitable couplings between them.
For sparse networks, the cliques are typically small and mostly
consist of two or three nodes.
This method thus enables the simulation of large networks much
beyond the point where the master equation becomes infeasible.

The moment equations include 
one equation for the population size of each reactive species
(represented by a first moment) 
and one equation for each reaction rate (represented by a second moment). 
These equations are obtained 
by taking the time derivative of each moment
and using 
the master equation
to express
the time derivatives of the probabilities
\cite{Lipshtat2003}.
In the resulting equations, 
the time derivative of each moment 
can be expressed as a linear combination of 
first, second and third order moments.
To close the set of moment equations 
one must express the third order moments in terms of
first and second order moments. 
This is achieved by the incorporatin of a suitable
truncation scheme of the master equation
\cite{Lipshtat2003,Barzel2007a,Barzel2007b}.
The moment equations provide the most efficient
incorporation of stochastic grain chemistry, including
hydrogen recombination and other 
surface reactions into models of interstellar
chemistry.

\section{Summary}
\label{sec:summary}

We have reported the results 
of experiments on molecular hydrogen formation 
on amorphous silicate surfaces, using 
temperature-programmed desorption.  
In these experiments
beams of H and D atoms were irradiated
on the surface of an amorphous silicate sample.
The desorption rate of HD molecules was monitored
using a mass spectrometer during a subsequent TPD run.
The results were analyzed using
rate equations. 
It was shown
that a model based on a 
single isotope provides the correct results for the activation 
energies for diffusion
and desorption of H atoms.
The barriers
$E_{\rm H}^{\rm diff}$,
$E_{\rm H}^{\rm des}$
and
$E_{\rm HD}^{\rm des}$,
as well as the density of adsorption sites
on the surface were obtained.

The results were used in order to evaluate the formation rate 
of H$_2$ on dust grains
under the actual conditions present in interstellar clouds. 
It was found that under typical conditions
in diffuse interstellar clouds, amorphous silicate grains
are efficient catalysts of H$_2$ formation when the grain
temperatures are between 9-14K. This temperature window is
within the typical range of grain temperatures in diffuse
clouds. It was thus concluded that amorphous silicates are
efficient catalysts of H$_2$ formation in diffuse clouds.
The recently developed computational methodologies for the
evaluation of reaction rates on interstellar grains
were briefly reviewed.

\acknowledgments
This work was supported 
by NASA through grants NAG5-9093, NAG5-11438
and NASA 1284471 (G.V),
by the Israel Science Foundation
and the Adler Foundation for Space Research (O.B), and 
by the Italian Ministry for University and Scientific Research
through grant 21043088 (V.P).

\clearpage
\newpage

\begin{table}
\caption{
Parameters for molecular hydrogen formation:
the diffusion and desorption barriers
of H atoms, the desorption barriers of HD molecules
and the density of adsorption sites.
}
\vspace{0.1in}
\begin{centering}
\begin{tabular}{lcccc}
\hline 
Material &
$E_{\rm H}^{\rm diff}$(meV) &
$E_{\rm H}^{\rm des}$(meV)  &
$E_{\rm HD}^{\rm des}$(meV) &
$s$ (sites cm$^{-2}$) \\
\hline 
Amorphous Silicate&
35&
44&
35, 53, 75 &
7 $\times 10^{14}$ \\
\hline
\end{tabular}
\par
\end{centering}
\label{t:E_barriers} 
\end{table}

\clearpage
\newpage

\clearpage
\newpage

\begin{thebibliography}{10}

\bibitem{Marenco1972}
{Marenco, G.; Schutte, A; Scoles, G.; Tommasini, J.}
Vac. Sci. Tech.
{\bf 1972}, {\it 9}, 824.

\bibitem{Schutte1976}
{Schutte, A; Bassi, D.; Tommasini, F.; Turelli, A.; 
Scoles, G.; Hermans, L.J.F.}
J. Chem. Phys.
{\bf 1976}, {\it 64}, 4135.

\bibitem{Govers1980}
{Govers, T.R.; Mattera, L.; Scoles, G.} 
J. Chem. Phys. 
{\bf 1980}, {\it 72}, 5446. 

\bibitem{Duley1984}
{Duley, W.W.; Williams, D.A.} 
{\em {Interstellar Chemistry}},
Academic Press, London, 1984.

\bibitem{Williams1998}
{Williams D.A.}, 
Faraday Discuss. Chem. Soc. 
{\bf 1998}, {\it 109}, 1

\bibitem{Gould1963}
{Gould, R.J.; Salpeter, E.E.} 
ApJ 
{\bf 1963}, {\it 138}, 393 

\bibitem{Hollenbach1970}
{Hollenbach, D.; Salpeter, E.E.} 
J. Chem. Phys. 
{\bf 1970}, {\it 53}, 79

\bibitem{Hollenbach1971a}
{D. Hollenbach and E.E. Salpeter}, 
ApJ 
{\bf 1971}, {\it 163}, 155

\bibitem{Hollenbach1971b}
{Hollenbach, D.; Werner, M.W.; Salpeter, E.E.} 
ApJ 
{\bf 1971}, {\it 163}, 165

\bibitem{Mathis1977}
{Mathis, J.S.; Rumpl, W.; Nordsieck, K.H.}
ApJ
{\bf 1977}, {\it 217}, 425

\bibitem{Draine1984}
{Draine, B.T.; Lee, H.M.}
ApJ
{\bf 1984}, {\it 285}, 89

\bibitem{Weingartner2001}
{Weingartner, J.C.; Draine, B.T.}
ApJ
{\bf 2001}, {\it 548}, 296

\bibitem{Tielens2005}
{Tielens, A.G.G.M.} 
{\em The Physics and Chemistry of the Interstellar Medium},
Cambridge University Press, Cambridge, 2005.

\bibitem{Hasegawa1992}
{Hasegawa, T.I.; Herbst, E.; Leung, C.M.}
ApJ Suppl.
{\bf 1992}, {\it 82}, 167.

\bibitem{Herbst2003}
{Herbst, E.; Shematovich, V.I.} 
Astrophys. Space Sci. 
{\bf 2003}, {\it 235}, 725.

\bibitem{Geppert2006}
{Geppert, W.D.; Hamberg, M.; Thomas, R.D.; \"{O}sterdahl, F.; Hellberg, F.;
Zhaunerchyk, V.; Ehlerding, A.; Millar, T.J.; Roberrts, H.; Semaniak, J.; 
af Ugglas, M.; K\"{a}llberg, A.; Simonsson, A.; Kaminska, M.; Lrsson, M.}
{Faraday Discussions} 
{\bf 2006}, {\it 133}, 177.

\bibitem{Hidaka2004}
{Hidaka, H.; Watanabe, N.; Shiraki, T.; Nagaoka, A.; Kouchi, A.} 
ApJ
{\bf 2004}, {\it 614}, 1124

\bibitem{Watanabe2005}
{Watanabe, N.}  
{\em IAU Symposium 231, Astrochemistry: Recent Successes
and Current Challenges}, 
eds. D.C. Lis, G.A. Blake and E. Herbst,
Cambridge University Press, Cambridge, UK, 
{\bf 2005}, 231, 415.

\bibitem{Garrod2006}
Garrod, R.; Park, I.H.; Caselli, P.; Herbst, E.
Faraday Discussions, 
{\bf 2006}, {\it 133}, 51.



\bibitem{Pirronello1997a}
{Pirronello, V.; Liu, C.; Shen, L.; Vidali, G.} 
ApJ 
{\bf 1997}, {\it 475}, L69.

\bibitem{Pirronello1997b}
{Pirronello, V.; Biham, O.; Liu, C.; Shen, L.; Vidali, G.} 
ApJ 
{\bf 1997}, {\it 483}, L131.

\bibitem{Pirronello1999}
{Pirronello, V.; Liu, C.; Roser J.E.; Vidali, G.} 
Astron. Astrophys. 
{\bf 1999}, {\it 344}, 681

\bibitem{Vidali1998}
{Vidali, G.; Pirronello, V.; Liu, C.; Shen, L.} 
Astrophys. Lett. and Commun. 
{\bf 1998}, {\it 35}, 423.

\bibitem{Rettner1994}
Rettner, C.T. 
J. Chem. Phys. 
{\bf 1994}, {\it 101}, 1529. 

\bibitem{Eilmsteiner1996}
Eilmsteiner, G.; Walkner, W.; Winkler, A. 
Surf. Sci. 
{\bf 1996}, {\it 352}, 263.

\bibitem{Khan2007}
Khan, A.R.; Takeo, A.; Ueno, S.; Inanaga, S.; Yamauchi, T; 
Narita, Y.; Tsurumaki, H; Namiki, A. 
Surf. Sci. 
{\bf 2007}, {\it 601}, 1635. 

\bibitem{Zecho2002}
Zecho, T.; Guttler, A.; Sha, X.; Lemoine, D.; Jackson, B.; Kuppers, J. 
Chem. Phys. Lett. 
{\bf 2002}, {\it 366}, 188.

\bibitem{Roser2002}
{Roser, J.E.; Manic\`o, G.; Pirronello, V.; Vidali, G.} 
ApJ 
{\bf 2002}, {\it 581}, 276.

\bibitem{Roser2003}
{Roser, J.E; Swords, S.; Vidali, G.; Manic\'o, G.; Pirronello, V.} 
ApJ
{\bf 2003}, {\it 596}, L55.

\bibitem{Hornekaer2003}
{Hornekaer, L.; Baurichter, A.; Petrunin, V.V.; Field, D.; Luntz, A.C.} 
Science
{\bf 2003}, {\it 302}, 1943.

\bibitem{Hornekaer2005}
{Hornekaer, L.; Baurichter, A.; Petrunin, V.V.; Luntz, A.C.; Kay B.D.; 
Al-Halabi, A.} 
J. Chem. Phys. 
{\bf 2005}, {\it 122}, 124701.

\bibitem{Dulieu2005}
{Dulieu, F.; Amiaud,  L.; Baouche, S.; Momeni, A.; 
Fillion, J.H.; Lemaire, J.L.} 
Chem. Phys. Lett.
{\bf 2005}, {\it 404}, 187.

\bibitem{Amiaud2006}
{Amiaud, L.; Fillion, J.H.; Baouche, S.; Dulieu, F.; 
Momeni, A.; Lemaire, J.L.}
J. Chem. Phys. 
{\bf 2006}, {\it 124}, 094702.

\bibitem{Creighan2006}
{Creighan, S.C.; Perry, J.S.A.; Price, S.D.} 
J. Chem. Phys. 
{\bf 2006}, {\it 124}, 114701. 

\bibitem{Williams2007}
{Williams, D.A.; Brown, W.A.; Price, S.D.; Rawlings, J.M.C.; Viti, S.} 
{Astron. Geophys.} 
{\bf 2007}, {\it 48}, 25.

\bibitem{Katz1999}
{Katz, N.; Furman, I.; Biham, O.; Pirronello, V.; Vidali, G.} 
ApJ 
{\bf 1999}, {\it 522}, 305.

\bibitem{Cazaux2002}
{Cazaux, S.; Tielens, A.G.G.M.}  
ApJ
{\bf 2002}, {\it 575}, L29.

\bibitem{Cazaux2004}
Cazaux, S.; Tielens, A.G.G.M.
ApJ
{\bf 2004}, {\it 604}, 222.

\bibitem{Perets2005}
{Perets, H.B.; Biham, O.; Pirronello, V.; Roser, J.E.; 
Swords, S.; Manic\'o, G.; Vidali, G.} 
ApJ 
{\bf 2005}, {\it 627}, 850.

\bibitem{Brucato2002}
{Brucato, J.R.; Mennella, V.; Colangeli, L.; Rotundi, A.; Palumbo, P.}
{Planet. Space Sci.} 
{\bf 2002}, {\it 50}, 829.

\bibitem{Perets2007}
{Perets, H.B.; Lederhendler, A.; Biham, O.; Vidali, G.; Li, L.; 
Swords, S.; Congiu, E.; Roser, J.E.; Manic\'o, G.; 
Brucato, J.R.; Pirronello, V.} 
ApJL
{\bf 2007}, {\it 661}, L163.

\bibitem{Vidali2006}
{Vidali, G.; Roser, J.E.; Li, L.; Congiu, E., 
Manic\'o, G.; Pirronello, V.}
{Faraday Discussions} 
{\bf 2006}, {\it 133}, 125.

\bibitem{Ghio1980}
Ghio, E.; Mattera, L.; Salvo, C.; Tommasini, F.; Valbusa, E. 
J. Chem. Phys. 
{\bf 1980}, 73, 556.

\bibitem{Cuppen2005}
{Cuppen, H.M.; Herbst, E.}
Mon. Not. R. Astron. Soc.
{\bf 2005}, {\it 361}, 565.

\bibitem{Biham2001}
{Biham, O.; Furman, I.; Pirronello, V.; Vidali, G.}
ApJ
{\bf 2001}, {\it 553}, 595.


\bibitem{Buch1991a}
Buch, V.; Czerminski, R.  
J. Chem. Phys.
{\bf 1991}, {\it 95}, 6026. 

\bibitem{Buch1991b}
Buch, V.; Zhang, Q. 
ApJ 
{\bf 1991}, {\it 379}, 647.

\bibitem{Vidali1991}
Vidali, G.; Ihm, G.; Kim, Y.-J.; Cole, M.W. 
Surf. Sci. Rep. 
{\bf 1991}, {\it 12}, 133

\bibitem{Biham1998}
{Biham, O.; Furman, I.; Katz, N.; Pirronello, V.; Vidali, G.} 
Mon. Not. R. Astron. Soc. 
{\bf 1998}, {\it 296}, 869

\bibitem{Biham2002}
{Biham, O.; Lipshtat, A.} 
Phys. Rev. E 
{\bf 2002}, {\it 66}, 056103.

\bibitem{Krug2003}
{Krug, J.;}, 
Phys. Rev. E 
{\bf 2003}, {\it 67}, 065012.

\bibitem{Lohmar2006}
{Lohmar, I.; Krug, J.} 
Mon. Not. R. Astron. Soc. 
{\bf 2006}, {\it 370},  1025

\bibitem{Caselli1998}
{Caselli, P.; Hasegawa, T.I.; Herbst, E.} 
ApJ 
{\bf 1998}, {\it 495},  309

\bibitem{Shalabiea1998}
{Shalabiea, O.M.; Caselli, P.; Herbst, E.} 
ApJ 
{\bf 1998}, {\it 502}, 652

\bibitem{Green2001}
{Green, N.J.B.; Toniazzo, T.; Pilling, M.J.; Ruffle, D.P.; 
Bell, N.; Hartquist, T.W.} 
Astron. Astrophys. 
{\bf 2001}, {\it 375}, 1111

\bibitem{Charnly2001}
{Charnley, S.B.} 
ApJ 
{\bf 2001}, {\it 562}, {L99}

\bibitem{Gillespie1976}
{Gillespie, D.T.} 
{J. Phys. Chem.} 
{\bf 1977}, {\it 81}, 2340

\bibitem{Stantcheva2002}
{Stantcheva, T.; Shematovich, V.I.; Herbst, E.} 
Astron. Astrophys. 
{\bf 2002}, {\it 391}, 1069

\bibitem{Stantcheva2003}
{Stantcheva, T.; Herbst, E.} 
Mon. Not. R. Astron. Soc. 
{\bf 2003}, {\it 340},  983

\bibitem{Lipshtat2004}
{Lipshtat, A.; Biham, O.} 
Phys. Rev. Lett. 
{\bf 2004}, {\it 93}, 170601 

\bibitem{Barzel2007c}
{Barzel, B.; Biham, O.; Kupferman R.}
Phys. Rev. E.
{\bf 2007}, {\it 76}, 026703.

\bibitem{Lipshtat2003}
{Lipshtat, A.; Biham, O.} 
Astron. Astrophys. 
{\bf 2003}, {\it 400}, 585

\bibitem{Barzel2007a}
{Barzel, B.; Biham, O.}
ApJ
{\bf 2007}, {\it 658}, L37

\bibitem{Barzel2007b}
{Barzel, B.; Biham, O.}
J. Chem. Phys.
{\bf 2007}, {\it 127}, 144703

\end{thebibliography}
\end{document}